\documentclass{midl} 

\usepackage{mwe}

\jmlrproceedings{MIDL 2020}{Medical Imaging with Deep Learning 2020}
\jmlrvolume{}
\jmlryear{}
\jmlrworkshop{MIDL 2020 -- Short Paper}
\editors{}

\title[Deep Learning For Head-Neck Outcome Prediction]{A Normalized Fully Convolutional Approach to Head and Neck Cancer Outcome Prediction}

\midlauthor{\Name{William Le\nametag{$^{1}$}} \Email{william.le@polymtl.ca}\\
\Name{Francisco Perdigon Romero\nametag{$^{1}$}} \Email{francisco.perdigon@polymtl.ca}\\
\Name{Samuel Kadoury\nametag{$^{1}$}} \Email{samuel.kadoury@polymtl.ca}\\
\addr Polytechnique Montréal}

\begin{document}
\maketitle

\begin{abstract}
In medical imaging, radiological scans of different modalities serve to enhance different sets of features for clinical diagnosis and treatment planning.
This variety enriches the source information that could be used for outcome prediction.
Deep learning methods are particularly well-suited for feature extraction from high-dimensional inputs such as images.
In this work, we apply a CNN classification network augmented with a FCN preprocessor sub-network to a public TCIA head and neck cancer dataset.
The training goal is survival prediction of radiotherapy cases based on pre-treatment FDG PET-CT scans, acquired across 4 different hospitals.
We show that the preprocessor sub-network in conjunction with aggregated residual connection leads to improvements over state-of-the-art results when combining both CT and PET input images.
\end{abstract}


\section{Introduction}
Cancer treatment planning remains a long process for the patient: from pre-treatment staging to post-therapy follow-up, many factors could have changed that can impact the effectiveness of the treatment.
One of the key decisions of the physician is the choice of line of therapy, for which automatic outcome prediction can be beneficial.
In head and neck cancer cases, positron emission tomography with fluorodeoxyglucose integrated with computed tomography (FDG PET-CT) for diagnosis and treatment planning \cite{castaldi2013role} can be used as inputs to deep learning-based medical image analysis models.

In previous works using this Cancer Imaging Archive (TCIA) Head-Neck-PET-CT dataset \cite{head-neck-pet-ct}, random forests were used to classify overall survival (OS) based on a combination of both PET and CT extracted radiomics features and clinical information \cite{Valli_res_2017}.
More recently, an end-to-end convolutional neural network (CNN) was used to successfully predict radiotherapy outcomes using only the planning CT scans as input using the same dataset \cite{Diamant_2019}.

In this work, we show that combining PET and CT image inputs improves binary classification performance.
For this purpose, a CNN architecture was implemented using residual connections \citet{resnet}, which have been shown to reduce vanishing gradient problems through identity shortcuts and thus allowing deeper models, and aggregated convolutions\citet{resnext}, which have been shown to reduce the number of parameters in a layer without loss of performance.
The combination of these two methods lead to a reduction in model size compared to previous state-of-the-art work by \citet{Diamant_2019}.
Furthermore, a fully-convolutional network (FCN) sub-network is used as a preprocessor.
\citet{Drozdzal_2018} showed previously that this technique acts as an image normalization on medical images for liver tumor segmentation.
We hypothesize that reusing this method for multi-modality inputs for classification, in combination with the aforementioned modifications, will lead to overall performance improvements when predicting patient survival from pre-treatment FDG PET-CT.

\section{Methodology and Experiments}
The Head-Neck-PET-CT\cite{head-neck-pet-ct} dataset used in this study consisted of 298 head and neck cancer patients acquired from 4 different institutions in Quebec.
Each patient had a pre-radiotherapy FDG PET-CT scan. Both PET and CT volumes were converted to 2D images using largest primary GTV lesion area slice selection.
Images were normalized to 0 mean and unit standard deviation.
The PET image was up-scaled to 512$\times$512 before being concatenated to the CT image to form the 512$\times$512$\times$2 input image.
The dataset presented only 56 (19\%) cases with the OS target (death).
Thus a resampling strategy with data augmentation (random flip, random shifts of 40\%, random rotations of 20 deg) was used to rebalance the dataset.

Our end-to-end binary classification model consists of two parts: a FCN sub-network \cite{Drozdzal_2018} and a fully convolutional classifier shown in Figure \ref{fig:architecture}. The FCN consists of 4 downsampling blocks followed by 4 upsampling blocks. Each block is composed of a 3x3 convolution layer with SeLU activation \cite{DBLP:journals/corr/KlambauerUMH17} and uses strided convolutions to modify the output dimension.  Output features of each downsampling block are concatenated with input features of each corresponding upsampling blocks.

\begin{figure}[!ht]
\floatconts
  {fig:architecture}
  {\caption{Proposed model architecture (top). The input consists of a 2 channel PET-CT image that is initially passed through a FCN (lower left). Downsampling uses convolutions with stride 2 while upsampling uses transposed convolutions. The output is then fed to an 18-layer deep CNN. Aggregated residual convolutional blocks (lower right) are repeated twice before being downsampled by setting the stride to 2. Classification is performed by taking the output vector with 256 features through a fully connected layer with softmax activation.}}
  {\includegraphics[width=\linewidth]{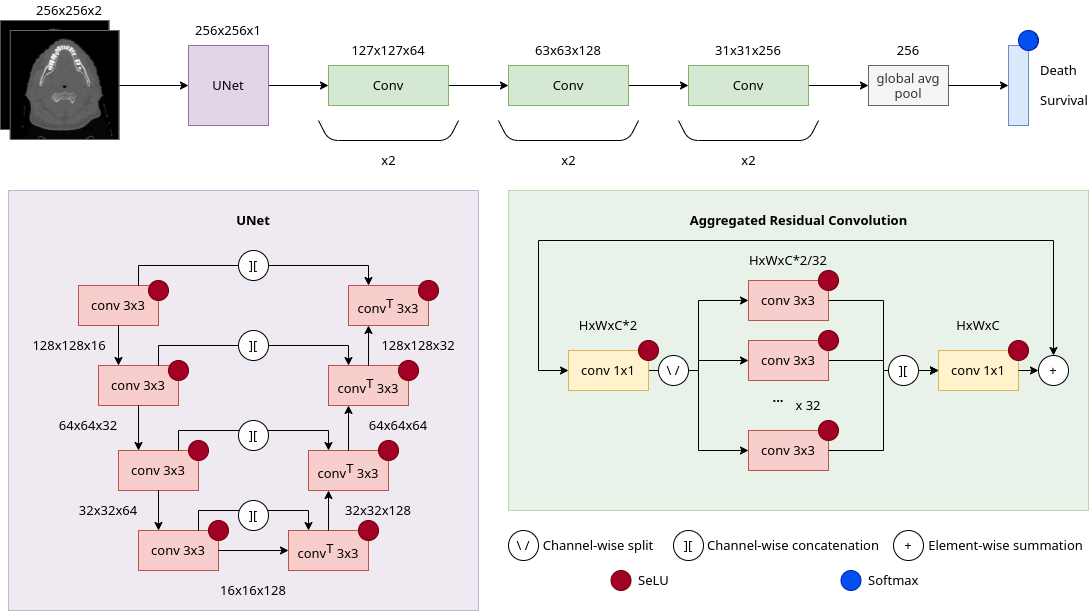}}
\end{figure}

The CNN classifier is inspired by the ResNeXt architecture\cite{resnext}, with 2 3x3 bottleneck layers with a filter growth factor of 2 and cardinality of 32 and residual connections around each. Downsampling is done using strided convolutions in the first layer of each block. Global average pooling \cite{resnext} and a fully connected layer are used to output the binary survival class.

Training is performed using categorical cross-entropy loss for classification (0: survival, 1: death) using the Adam optimizer \cite{kingma2014adam} with a learning rate of 0.0006 for 100 epochs and a batch size of 8. No early-stopping techniques were used. The code is implemented in Keras\cite{chollet2015keras} and trained on a GeForce RTX 2080 Ti for 1 hour.

\section{Results and Conclusions}
Binary classification results are monitored using the area under the receiver-operating characteristic curve (AUC). Table \ref{tab:performance} shows the progression of modifications from the CNN as presented by \citet{Diamant_2019} up to our proposed model, using the different input modality.

The best performing model using the CNN architecture proposed by \citet{Diamant_2019} used the masked CT input which is the same model input as was used in their report. The performance observed is lower (0.67 $<$ 0.70) due to different initial seed conditions. With the addition of the FCN preprocessor, the performance drops to 0.59 for PET, 0.65 for CT, 0.63 for masked CT but improves to 0.70 for PET+CT. This shows that the FCN works by mapping features from multiple domains (modalities) to a latent embedding that benefits classification performance.

Our proposed CNN with aggregated convolutions and residual connections (AggResCNN) performs better than previous state-of-the-art deep learning methods (0.74 $>$ 0.70) and previous radiomics-based random forest methods (0.74 $>$ 0.65) when considering the area under the receiver-operating characteristic curve (AUC) and using both PET and CT as inputs
With the addition of the FCN preprocessor, this performance improves to 0.76 AUC.
The final proposed FCN+AggResCNN thus improves over previous methods (0.76 $>$ 0.70 $>$ 0.65) on the same dataset while having less total parameters (683650 $<$ 930146).

Furthermore, the model trained on masked CT performs better than the whole CT for the CNN case (0.67 $>$ 0.57) while the reverse is true for the FCN+CNN case (0.63 $<$ 0.65).
Similarly, the masked CT performance is greater than the full CT image for the AggResCNN (0.69 $>$ 0.65), while the reverse is true for the FCN+AggResCNN (0.67 $<$ 0.70). This suggests that the FCN learns features on information outside the GTV boundaries to be effective.

In every case, classification performance on just the PET input is worse than the rest, which suggests that this imaging modality only contributes auxiliary information that complements the CT image. The cause of this apparent worse performance can be attributed both to the lower image resolution and the absence of image information in PET away from the tumor lesion, as healthy tissues don't absorb the radioactive tracing agent.
\begin{table}[htbp]
\floatconts
  {tab:performance}
  {\caption{Classification performance of the proposed models compared to state of the art results on the Head and Neck FDG PET-CT TCIA Dataset. The first model consists of a random forest for radiomics features selection followed by another random forest for classification. The second model consists of a CNN with 3 convolutions/PReLU/max-pooling layers and 2 fully connected layers. The last three models show ablation results of our proposed FCN 18-ResNeXt trained on PET and CT images.}}
  {\begin{tabular}{llccccc}
  Model & Datasets & Parameters & AUC & Spec & Sens \\
  \hline
  \citet{Valli_res_2017} & PET+CT + clinical & - & 0.65 & 67\% & 55\% \\
  \citet{Diamant_2019} & CT masked & 930146 & 0.70 & 67\% & 68\% \\
  CNN & PET & 930146 & 0.59  & 90\% & 29\% \\
  CNN & CT & 930146 & 0.57 & 37\% & 77\% \\
  CNN & masked CT & 930146 & 0.67 & 82\% & 52\% \\
  CNN & PET + CT & 930946 & 0.65 & 99\% & 30\% \\
  FCN + CNN & PET & 1321682 & 0.59 & 41\% & 77\% \\
  FCN + CNN & CT & 1321682 & 0.65 & 51\% & 79\% \\
  FCN + CNN & masked CT & 1321682 & 0.63 & 35\% & 90\% \\
  FCN + CNN &  PET + CT & 1322482 & 0.70 & 69\% & 71\% \\
  AggResCNN & PET & 291874 & 0.50 & 100\% & 0\% \\
  AggResCNN & CT & 291874 & 0.65 & 54\% & 76\% \\
  AggResCNN & masked CT & 291874 & 0.69 & 51\% & 87\% \\
  AggResCNN & PET + CT & 292114 & 0.74 & 66\% & 82\% \\
  FCN+AggResCNN & PET & 683410 & 0.57 & 21\% & 94\% \\
  FCN+AggResCNN & CT & 683410 & 0.70 & 46\% & 94\% \\
  FCN+AggResCNN & masked CT & 683410 & 0.67 & 52\% & 82\% \\
  FCN+AggResCNN & PET + CT & 683650 & 0.76 & 61\% & 91\% \\
  \end{tabular}}
\end{table}
Finally, the proposed architecture using aggregated convolutions and residual connections improves over state-of-the-art deep learning methods and radiomics based machine learning method by incorporating additional imaging data in the form of PET images. With the adapted architecture, the model can extract more features that help binary survival outcome prediction all while lowering the memory cost of the model using state-of-the-art convolutional techniques.

Thus, after training on both CT and PET images, our proposed model has overall less total parameters (683 650 $<$ 930146) and improves in AUC by 6 percentage points over the state of the art.

\bibliography{bibliography}
\end{document}